\documentclass[11pt,a4paper]{article}
\usepackage[utf8]{inputenc}
\usepackage[T1]{fontenc}
\usepackage{lmodern}
\usepackage[margin=2.5cm]{geometry}
\usepackage{amsmath,amssymb}
\usepackage{graphicx}
\usepackage{setspace}
\usepackage[font=small,labelfont=bf]{caption}
\usepackage[hidelinks]{hyperref}
\usepackage{xurl}
\usepackage{xcolor}

\graphicspath{{figs/}}
\hypersetup{
  pdftitle={Characterizing landslide dynamics from time-lapse time domain induced polarization and ground-based imaging: a case study of the MontGombert landslide (French Alps)},
  pdfauthor={A. Carrier, O. Meric, P. Bottelin},
  pdfsubject={Preprint},
  pdfkeywords={Landslides, Electrical tomography, Induced polarization, Permeability, Surface displacement, Risk assessment}
}

\newcommand{\relvar}{\dfrac{\text{variable} - \text{variable}_{ref}}{\text{variable}_{ref}}}

\begin{document}

\begin{center}
\fbox{\parbox{0.92\textwidth}{\small\setstretch{1.1}
\textbf{Preprint.} This is the authors' submitted version of the article published in
\emph{Landslides}, 21 (2024), 353--369.
The final version is available at \url{https://doi.org/10.1007/s10346-023-02137-0}.
This version may differ from the published version.}}
\end{center}

\vspace{1em}
\begin{center}
{\Large\bfseries Characterizing landslide dynamics from time-lapse time domain induced polarization and ground-based imaging: a case study of the MontGombert landslide (French, Alps)\par}
\vspace{1.2em}
{\large Aurore Carrier$^{*}$, Ombeline Meric, Pierre Bottelin\par}
\vspace{0.8em}
{\small\setstretch{1.1}
ADRGT, SAGE, Rue de la Condamine, 38610 Gières, Isère, France\\
$^{*}$Corresponding author: a.carrier@sage-ingenierie.com\par}
\end{center}

\section*{Abstract}
Electrical resistivity tomography (ERT) is a common tool to study landslide morphology and dynamics. Landslide dynamics is strongly related to clay content, pore pressure and permeability distribution. The standard direct current (DC) resistivity method often fails at providing information about permeabilities because materials with low resistivities can both exhibit high or low permeability especially according to their clay content and the relative importance of surface conduction compared to electrolytic conduction. Induced polarization (IP) has proven to be a useful complementary tool to assess petrophysical properties of the rocks. Up to now, it has been rarely used in landslide studies and not yet in landslide monitoring. This study aims at demonstrating the usefulness and added value of IP as a complement to traditional DC measurements in the case of landslide characterization and monitoring. We performed 3 time domain IP (TDIP) surveys at three different time steps on an active landslide in the French Alps. A time-lapse inversion of the resistivity and chargeability is realized. The spectral IP parameters are extracted through a Debye decomposition. The results are compared to surface displacements obtained by processing optical images. We demonstrated that the use of TDIP is of crucial importance to estimate the extent of unstable volumes within the landslide. The results are in very good agreement with the data provided by the piezometers and inclinometers. The use of a general law to link IP variables and permeability enables to estimate the spatial distribution of the permeability within the landslide and its evolution through the seasons.

\medskip
\noindent\textbf{Keywords:} Landslides $\cdot$ Electrical tomography $\cdot$ Induced polarization $\cdot$ Permeability $\cdot$ Surface displacement $\cdot$ Risk assessment

\section{Introduction}

Landslide occurrences are increasing nowadays because climate is changing and society and land use are evolving and extending. Understanding landslide processes and the associated risk is becoming a major economic and political stake, especially in mountainous areas (Malet et al. 2007; Marescot et al. 2008; Gance et al. 2016; Lajaunie et al. 2019; Scoppettuolo et al. 2020; Desrues et al. 2022). Landslide deformation monitoring is one type of approach to evaluate slope stability and set up early warning techniques. It implies surface deformation monitoring with satellite or image-based approaches and/or deformation modeling approaches (Stumpf et al. 2017; Lacroix et al. 2018; Desrues et al. 2019; Desrues et al. 2022; Cascini et al. 2022).

Full landslides characterization consists in (i) evaluating the potential unstable volumes and (ii) defining the triggering processes. Temporal changes in water content of the subsurface material, the associated pore water pressure and matrix suction variations are the main involved triggering parameters. It underlies the strong influence of heavy rainfalls and sudden snow melts in landslide dynamics (Malet et al. 2005; Eberhardt et al. 2005; Guzzetti et al. 2008; Lu et al. 2010; Ponziani et al. 2012; Yao et al. 2019). Geophysical approaches, and especially electrical methods, due to their sensitivity to water content, have proven to be efficient for structure characterization, soil moisture content characterization and its evolution through time (Marescot et al. 2008; Hibert et al. 2012; Wilkinson et al. 2016; Gance et al. 2016; Hellman et al. 2017; Uhlemann et al. 2017; Lajaunie et al. 2019; Revil et al. 2020; Holmes et al. 2020). If moisture content is a main factor controlling landslide stability assessment, the distribution of clayish areas acting as impermeable barriers is also a key to understand landslides dynamics. Recently, Uhlemann et al. (2017); Chambers et al. (2014) proposed a moisture content evaluation based on the measure of the electrical resistivity and the consideration of the surface conduction, based on the laboratory calibration of a Waxman and Smith type relationship.

Considering the importance of clay content estimation and distribution in understanding landslides, it turns out logical to introduce induced polarization (IP) measurements to classical geoelectrical surveys. IP measurements can be performed both in frequency (FDIP) and in time (TDIP) domains. Both domains can yield information about the amplitude of the polarization and the relaxation time. The IP spectral content holds information about petrophysical and hydrological characteristics such as the hydraulic conductivity (Binley and Kemna 2005; Tong et al. 2006a; Titov et al. 2010; Kemna et al. 2012) or the clay content (Slater and Lesmes 2002; Slater and Glaser 2003). Up to now, very few studies present the results of field-induced polarization on landslides (Marescot et al. 2008; Revil et al. 2020; Orozco et al. 2022). Revil et al. (2020) applied petrophysical models in order to estimate the water content, cation exchange capacity and permeability from laboratory calibration to a clayey landslide. Similarly to this study, most of the field IP and almost all the field IP realized on landslides consider only the chargeability parameter depriving themselves of the spectral information. Recently, Orozco et al. (2022) have performed both TDIP and FDIP on a landslide and have evidenced that hydraulic properties estimated from spectral IP content were of fundamental importance for landslide characterization. IP spectral content is usually retrieved from FDIP measurements due to their higher accuracy and larger acquisition bandwidth. Such acquisitions are yet time greedy and have a lower penetration depth on the field (Maurya et al. 2018; Martin et al. 2020). In the past 10 years, the use of TDIP to retrieve IP spectral information has been developing. Generally, the spectral properties are extracted via a Cole Cole, a Pelton or a maximum phase angle model inversion (Gazoty et al. 2012; Fiandaca et al. 2013; Doetsch et al. 2015; Fiandaca et al. 2018; Lévy et al. 2019; Martin et al. 2020; Orozco et al. 2022). The latter imply some hypothesis as for the shape of the phase spectra. To avoid such constraints, TDIP decays can also be fitted via a Debye decomposition (DD) (Tarasov and Titov 2007; Nordsiek and Weller 2008; Martin et al. 2020, 2021). To our knowledge, the DD approach has never been applied on landslide, and the spectral content evolution through time in landslides has not yet been investigated.

In this paper, we propose a time-lapse ERT and IP monitoring of a landslide situated in the French Alps. The full decay of the IP curve will be considered in order to get information about the relaxation times via a Debye decomposition. We propose to associate the results to the surface deformation obtained by photogrammetric monitoring. We show that the IP information is important as an added value for landslides geometry characterization, time-lapse monitoring, dynamic processes understanding and unstable volumes quantification.

\section{Geological settings and major stakes}
\label{sec:geol}

\begin{figure}[htbp]
\centering
\includegraphics[width=\textwidth]{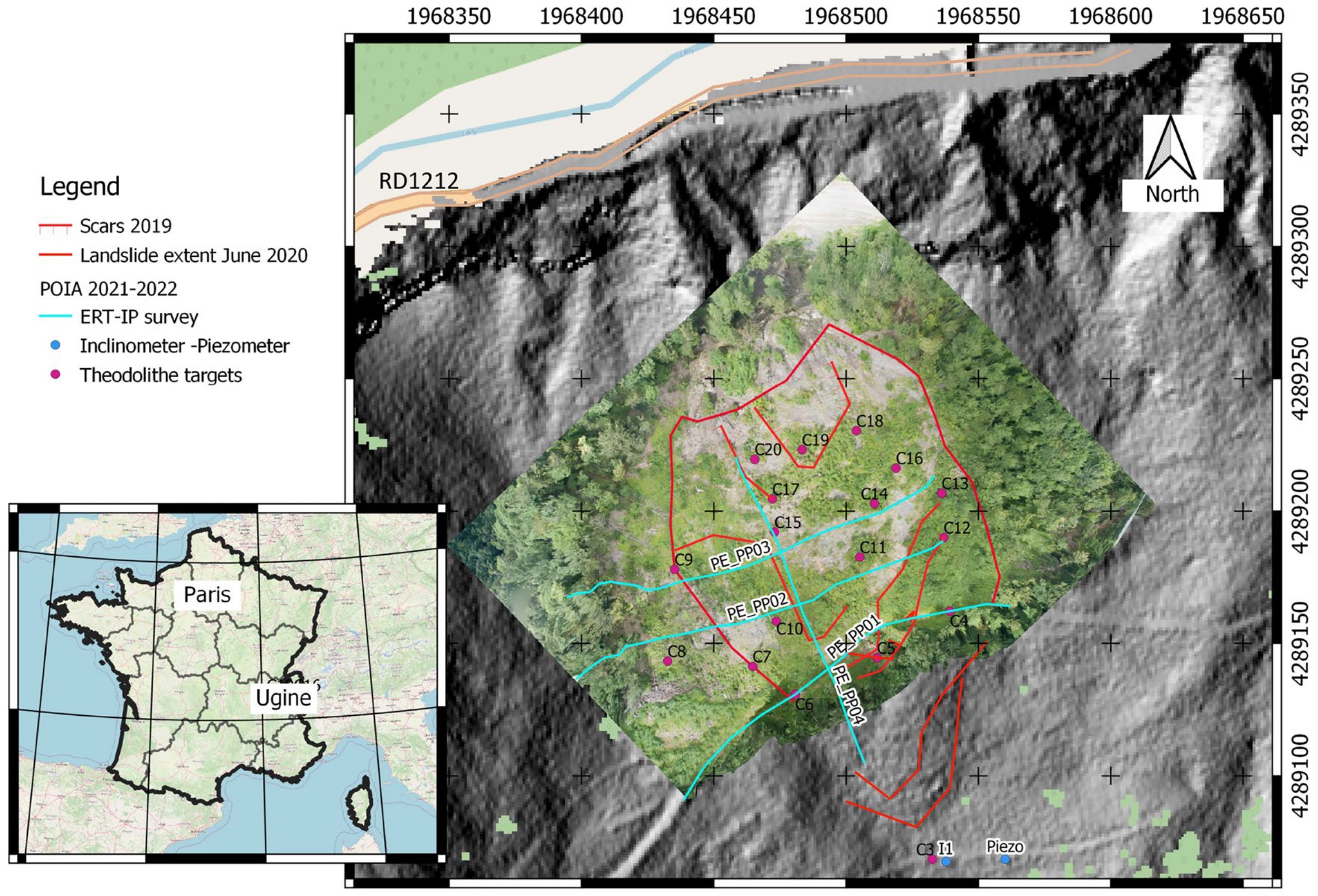}
\caption{Location of the studied area. (Right panel) Map of the Montgombert landslide (MGL). The spatial extent of the most active part of the landslide is represented by red lines, and the mass is sliding towards the North. The location of the geophysical surveys (blue lines), inclinometers-piezometer (blue points) and theodolite targets (pink points) are also represented. Coordinates are given in Lambert CC zone 45.}
\label{fig:loc}
\end{figure}

The Montgombert landslide (MGL) is located in the French part of the alps in the gorges of the Arly river (Savoie, France) near the city of Ugine (Fig.~\ref{fig:loc}). The landslide is perched 100~m above the road RD1212. This road is an essential trunk road as it serves some major ski resorts in the area and connects cities in the Pays du Mont Blanc to bigger cities. The part of the RD1212 crossing the gorges of the Arly has been regularly exposed to landslides, rock falls and flooding, generating regular traffic interruptions and road works. All the Eastern part of the gorges de l'Arly are prone to landslides and are composed of Hercynian metamorphic micaschists from the Série Satinée. The MGL originates in the generalized toppling failure of the micaschists and their transformation into clayey micaschists rock slides. Further on the East, the micaschists are covered by mesozoic rocks and glacial sediments on the Plateau des Saisies. The latter forms significant local aquifers that generate permanent water circulation into the massif. Water is drained by the surface hydrological network and also diffusively through the rock slides. The MGL has been particularly active since the spring 2016 and strongly monitored ever since. Piezometers and inclinometers have been installed. Surface displacements are monitored thanks to automatic theodolite and optical image monitoring installations (Desrues et al. 2019). The slope is the order of 35\textdegree. Geophysical studies have been performed in 2016 and 2017. The most active part of the landslide is about 100~m width and 200~m long, and its depth is evaluated to 15 to 30~m. The piezometric level at the top of the active part of the landslide is 18 to 16~m below the topographic surface in spring and 30 to 31~m at the end of the summer. The surface displacements show that movements are more important in the lower part of the landslide and can reach 25~m within a year. The inclinometer exhibits 3 main sliding surfaces at 8, 15 and 30~m depth, the latter being consistent with the water level observed in the piezometer. The landslide is water sensitive and is particularly active in the spring following heavy rains and/or fast snow melting. The particularly rainy winter 2017--2018 generated a sudden reactivation of the MGL. Open fractures up to 2~m wide and 10~m long have been produced, 7000~m$^3$ of rock debris have reached the road, and several hundreds of thousands cubic meters of material have moved from 50~cm to 15~m at the surface of the landslide and stayed in equilibrium on the slope. As a consequence, the administration of the French county of Savoie decided to finance more geological, hydrogeological and geophysical studies to better understand the structure of the landslide and the mechanisms driving its instabilities.

\section{Data acquisition and pre-processing}

\subsection{Photogrammetric data}

A system of topographic monitoring is in place on the studied site (Fig.~\ref{fig:loc}). We placed 26 topographic targets in 2016 and surveyed their positions with an automatic total station. Following technical issues and some target loss, the installation had to be reset and has been operational from February 2021 to November 2021 and further removed. We installed 2 digital single-lens reflex cameras (Canon 100D) of 18~MPix connected to a Paratronic LNS datalogger and separated of 362~m on the northern opposite side (right bank) of the valley enabling a full view of the most active part of the landslide. The cameras have been installed in May 2019 and are operating ever since. The distance between the cameras and the landslide is 550~m, the equivalent size of one pixel is 5~cm, and the theoretical displacement detection limit is 0.5~cm (Provost et al. 2022). In order to avoid problems related to shades, 3 pictures per day are taken at 11:00, 12:00 and 13:00 (UTC). The calibration of the cameras has been performed thanks to 10 georeferenced targets placed on the landslide and a high-resolution Lidar obtained at the moment of the installation of the cameras. We used the images from the 18th of April 2021, the 16th of September 2021 and the 12th of April 2022. Each pair of images has been processed.

\subsection{Geoelectrical data}

We acquired 4 electrical ERT and IP profiles in the active part of the MGL. A 5~m spacing multiple gradient acquisition protocol with a double cable configuration (2 $\times$ 32 stainless steel electrodes) has been performed in order to maximize signal/noise ratio and prevent electromagnetic coupling effects within the cables (further details about this acquisition procedure can be found in Dahlin et al. (2002) and Lévy et al. (2019)). We realized 3 campaigns in April 2021, September 2021 and May 2022. The studied area being particularly steep and unstable, the electrodes have not been left in place from one campaign to the other. At each campaign, the precise position of each electrode is taken with a real-time kinematic GPS (Leica GS07 high precision GNSS with a RTK connection) with a 5~cm accuracy.

In April 2021, resistivity and induced polarization data were acquired using an ABEM Terrameter SAS 4000. For IP data, 6 time windows (100, 200, 400, 800, 1600, 3200~ms) were set with an initial delay of 50~ms to make sure no electromagnetic coupling is recorded on the early times; on-time and off-time were equal and set to 8~s. To reduce the acquisition time in September 2021 and May 2022, a Terrameter LS2 has been used in 100\% duty cycle (Olsson et al. 2015). The full-waveform data enabled a first data quality check, manual inspection and removal of noisy data (Martin et al. 2020). Injection time was set to 8~s, and 13 time windows have been recorded with an initial delay of 50~ms (100, 100, 100, 100, 100, 100, 100, 100, 200, 400, 800, 1600, 3200~ms). Unfortunately, the chargeability values are shifted from 50\% duty cycle to 100\% duty cycle, and we cannot ensure the comparability of these data and their decays between April 2021 and the further campaigns (Olsson et al. 2015). This is why we will not further use the April 2021 chargeability data for TDIP time-lapse inversion. A total of 1053, 996 and 977 resistivity and chargeability raw data have been obtained in April 2021, September 2021 and May 2022, respectively. We measured the stacking errors---or variation coefficient---in \% over at least 4 stacks. The median value and standard deviation value of these errors are 0.72 and 1.55\% in April 2021, 0.89 and 0.17\% in September 2021 and 0.085 and 0.12\% in May 2022. The errors are 10 times less spread when using the LS2 which can be explained by a more recent manufacture and a better optimized internal quality. The stacking errors in September are higher than those obtained in May because of higher contact resistances at the end of the summer due to dry superficial soil layer (1000~Ohm with a maximum of 1700 in September and 700~Ohm in May). Negative resistivities, values with stacking errors superior to 2\%, 1.2\% and 1\% for April 2021, September 2021 and May 2022 as well as outliers values are removed from the resistivity data set. Chargeability decays are checked and filtered such as negative chargeabilities at first step; decays which do not fit an exponential decay and non-monotonic decreasing curves are removed. A total of 1028, 895 and 969 resistivity data and 594, 747 and 905 chargeability decays are kept for further inversion (April 2021, September 2021 and May 2022, respectively).

\section{Methods}

\subsection{Electrical and induced polarization methods}

Classical ERT measurements consist in injecting current through a pair of electrodes and measuring the resulting voltage through a pair of receiving electrodes leading to a resistance value. However, the subsurface materials can also act as capacitors, i.e., they can reversibly store electrical charges when an external current is applied. This effect is also known as induced polarization effect or chargeability. Charges accumulation and depletion generate a secondary electrical field. In time domain (TD), this leads to a voltage decay when the current is switched off. The characteristics of the decay curve are fundamental to define the IP effect not only in terms of magnitude but also in terms of relaxation time (Gazoty et al. 2012; Martin et al. 2021). The overall conductivity of soils (inverse of resistivity) is the result of three main effects: the matrix conduction, which can be neglected if no metallic particles or electronic conductors are present; the electrolytic conduction, i.e., the conduction of the charges within the fluid present in the pores; and the surface conduction, i.e., conduction processes happening at the fluid/grain interface. In materials characterized by fine grains like clays, surface conduction cannot be neglected. When no metallic particles are present, the conductivity is a complex value that can be expressed as:
\begin{equation}
\sigma^* = \left[\sigma_{el} + \sigma_s'\right] + i\sigma_s''
\label{eq:sigma}
\end{equation}
with $\sigma_{el}$ the electrolytic conductivity and $\sigma_s$ the surface conductivity, both in S.m$^{-1}$. Capacitive properties are related to the imaginary part of the surface conductivity (Kemna et al. 2012). The normalized chargeability is defined as $m_n = \frac{m}{\rho_0}$, where $m$ is the chargeability and $\rho_0$ the measured DC resistivity. This variable is considered as analogical to the imaginary part of the conductivity, $\sigma''$, in the frequency domain (FD) (Slater and Lesmes 2002; Gurin et al. 2013) and can be considered as a proxy for surface conductivity in sedimentary rocks (Slater and Lesmes 2002; Revil et al. 2017). Practically, a sequence of pulses of constant current and opposite signs with or without pauses between them (respectively called 50\% versus 100\% duty cycles (Olsson et al. 2015; Madsen et al. 2017)) are injected into the ground. Partial apparent chargeabilities, in V/V, are defined as (Slater and Lesmes 2002; Gazoty et al. 2012; Revil et al. 2020)
\begin{equation}
m_i = \frac{1}{\left(t_{i+1} - t_i\right)} \int_{t_i}^{t_{i+1}} \eta(t)\,dt
\label{eq:mi}
\end{equation}
where $\eta(t) = \frac{V_s(t)}{V_{DC}}$ in V/V is the ratio between the secondary potential measured during the off-time, $V_s(t)$, and the potential measured between the potential electrodes just before the current is switched off, $V_{DC}$, and $t_{i+1}$ and $t_i$ define the beginning and ending of the window over which the signal is integrated.

\subsection{Geoelectrical inversion including induced polarization data}

The inversion of the electrical and TDIP data has been performed thanks to the open-source package pyGIMLI (Günther et al. 2006; Rücker et al. 2006; Rücker et al. 2017). As a first step, the inversion of the resistivity data is performed using a Gauss-Newton minimization algorithm implemented on an irregular grid (Günther et al. 2006; Rücker et al. 2006). A Gaussian distribution of the error estimate on the data is considered. The errors are defined as the sum of the measurement accuracy, resulting from the measurement of stacking errors (Tso et al. 2017) and a fixed percentage of the measure (Günther et al. 2006). The stopping criterion is based firstly on the $\chi^2$ criterion and then on the decrease of the cost function. The model resolution is estimated via the coverage approach developed in Günther et al. (2006). Considering the approach of Seigel (1959), the apparent chargeabilities are then inverted to get a chargeability model (Martin et al. 2020; Revil et al. 2020). One model per time window is extracted, and a smoothing constraint is applied along the time axis to ensure the modeled decays to be smooth (Martin et al. 2020). The resulting decays can further be studied.

\subsection{Extract information from the decay curves through a Debye decomposition}

Commonly, the inversion of TDIP uses the DC resistivity and the integral chargeability (the partial chargeability integrated over all the time gates) to obtain resistivity and chargeability models (Oldenburg and Li 1994; Revil et al. 2020). However, no information about the IP spectral content can be extracted this way. The Debye decomposition (DD) approach has been used in frequency domain by Nordsiek and Weller (2008) and in time domain by Tong et al. (2006a), Tarasov and Titov (2007) and Martin et al. (2021). In this paper, we follow the approach proposed by Tarasov and Titov (2007) and more recently by Martin et al. (2021) in time domain. The decay curves are converted into an equivalent sum of Debye models via a Relaxation Time Distribution (RTD). The relaxation times are logarithmically discretized with 200 steps in a predefined matrix from $10^{-3.5}$ to $10^{3.5}$~s. The RMS and the $\chi^2$ values are extracted to ensure the RTD inversion validity. We also define the mean relaxation time $\tau_{mean}$ as (Nordsiek and Weller 2008)
\begin{equation}
\tau_{mean} = \exp\left(\frac{\sum_k m_k \ln(\tau_k)}{\sum_k m_k}\right)
\label{eq:tau}
\end{equation}

\subsection{Time-lapse inversion methodology}

To track the evolution and variations of resistivity and chargeability values along time and infer the influence of the water content in the MGL, the ERT and IP experiments are also conducted as a time-lapse experiment. The accessibility and the stability of the studied area remained challenging. Some areas at the base of the studied site became too unstable from one campaign to the other to ensure a safe access implying some electrodes relocation. As a result, our methodology for the time-lapse inversion is similar to the methodology proposed by Uhlemann et al. (2017). A very precise mesh is built on a 20~cm z-precision digital elevation model (DEM) and includes one node per electrode position all campaigns combined. The electrode positions are updated for each time step, and a reference model is used to constrain the inversion. The same inversion mesh is used for all the inversions as no significant topographic changes compared to the grid size that may affect the resistivity model are observed during our time study. Above 0\,\textdegree C, conductivity and normalized chargeability increase linearly of about 2\% per \textdegree C (Hayley et al. 2007; Revil et al. 2017; Uhlemann et al. 2017; Holmes et al. 2020). In order to avoid misinterpretation about water content variation, the influence of the seasonal temperature needs to be removed. Brunet et al. (2010) proposed to model the temperature as a function of depth, annual mean temperature, yearly amplitude variation of the air temperature, a characteristic penetration depth and a phase offset. This has been successfully applied by Chambers et al. (2014), Uhlemann et al. (2017) and Holmes et al. (2020). We obtained the parameters of the temperature model using field data retrieved from temperature sensors set at 40 and 85~cm depth at the studied site and meteorological data recorded within a year. Resistivity values can then be corrected for the effect of the temperature considering:
\begin{equation}
\rho_{corrected} = \rho\left(1 + \alpha\left(T_{ref} - T_{model}\right)\right)
\label{eq:temp}
\end{equation}
where $\rho_{corrected}$ is the resistivity corrected from the effect of the temperature, $\rho$ is the initial considered resistivity point, $T_{ref}$ is the reference temperature, and $T_{model}$ is the estimated temperature at the time and depth of the resistivity point. We set $\alpha$ for resistivity and normalized chargeability to 0.2 and 0.21 according to the values commonly used in the bibliography and obtained by laboratory experiments (Uhlemann et al. 2017; Coperey et al. 2019; Holmes et al. 2020).

\subsection{Clustering}

Data interpretation is often biased, because it is dependent upon the operator, especially as for what a significant variation is. In recent near surface geophysics applications, unsupervised classification has proven to be an efficient tool to define groups of common geophysical characteristics (Ronczka et al. 2017; Whiteley et al. 2021; Carrier et al. 2022) or to identify time-lapse variations (Xu et al. 2017; Delforge et al. 2021). We apply unsupervised classification to the variations of the different geophysical properties in order to create groups areas where geophysical properties evolve through time in a similar manner. For our specific case, we choose to apply a Gaussian mixture model (GMM) for its ability to identify convex clusters in the parameter space and for its speed and easiness of implementation via the scikit-learn python package (Pedregosa et al. 2011). The GMM clustering algorithm is performed to build consistent groups of (resistivity-normalized chargeability-relaxation time) variations. The data space consists of the 70\% of data with the highest coverage values; we fixed the number of groups to 5 considering the geological context, the rationale and a BIC (Bayesian information criterion) approach.

\subsection{Surface displacements from ground-based images (TSM)}

Thanks to the recent development of high-resolution sensors, the use of optical cameras to track surface displacements via image correlations appears to be relevant for landslide monitoring (Gance et al. 2014; Travelletti et al. 2012). These methods have the advantage of being low cost and providing information within the whole image and not only at specific points. We used the TSM (Tracing Surface Motion) processing toolbox developed by Desrues et al. (2019) and the VSC (Variable Sequential Correlation) methodology to compute the deformation fields. It consists of a 6-step methodology based on the use of a time series of images in a jpeg format originated from a monoscopic camera. The 6 steps are detailed in Desrues et al. (2019) and include evaluating the sensor orientation, selecting images, correcting the sensor movements, cross-correlating images to get ground motion, extracting the meaningful patterns and applying a geometrical correction.

\section{Results}
\label{sec:results}

\begin{figure}[htbp]
\centering
\includegraphics[width=\textwidth]{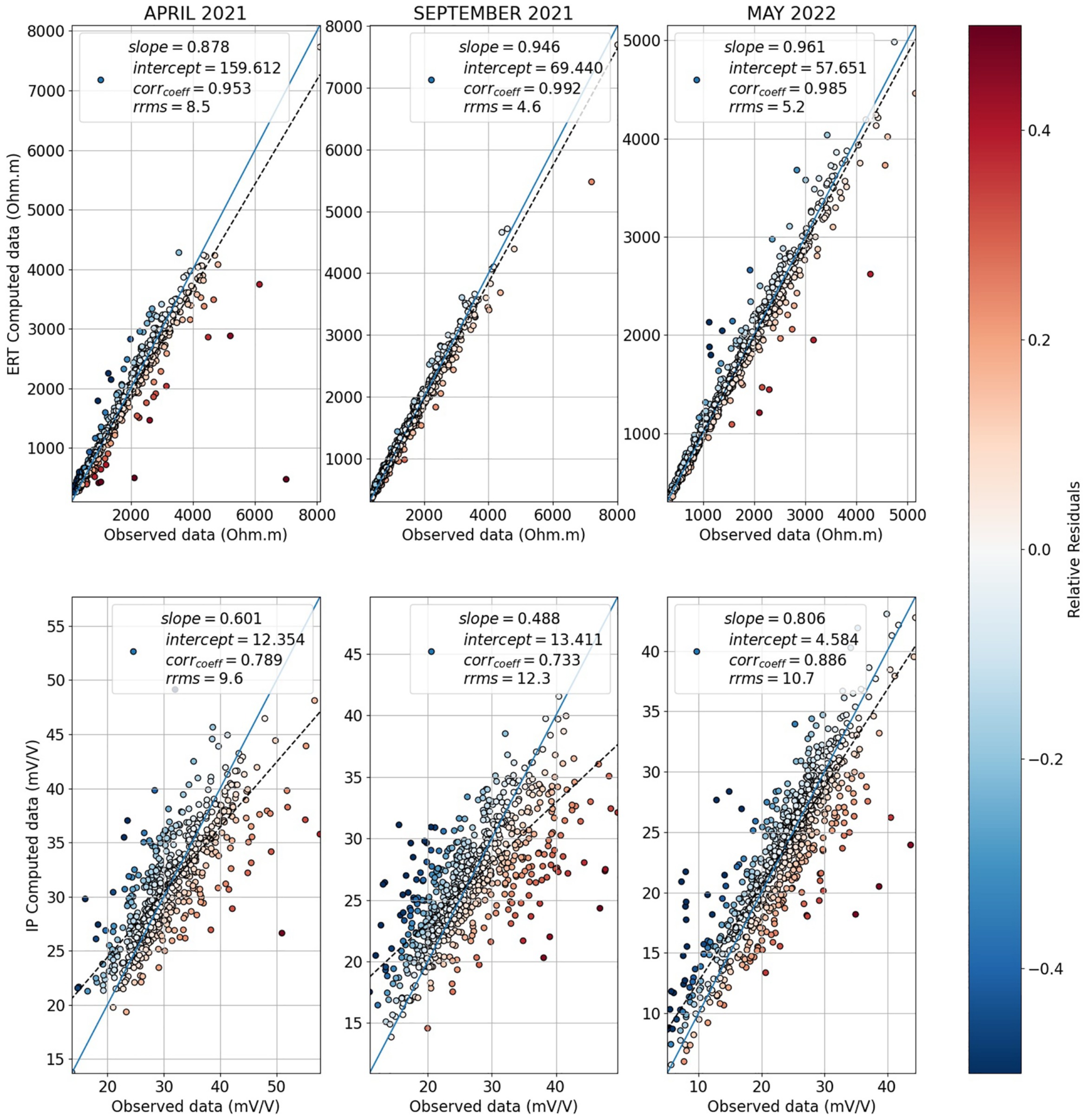}
\caption{Cross-plots obtained from the ERT and IP time-lapse inversion. The color scales with the relative residuals. The blue line is $y = x$ line, and the dotted black line is the least-square regression curve of the computed versus observed data points.}
\label{fig:xplot}
\end{figure}

\begin{figure}[htbp]
\centering
\includegraphics[width=\textwidth]{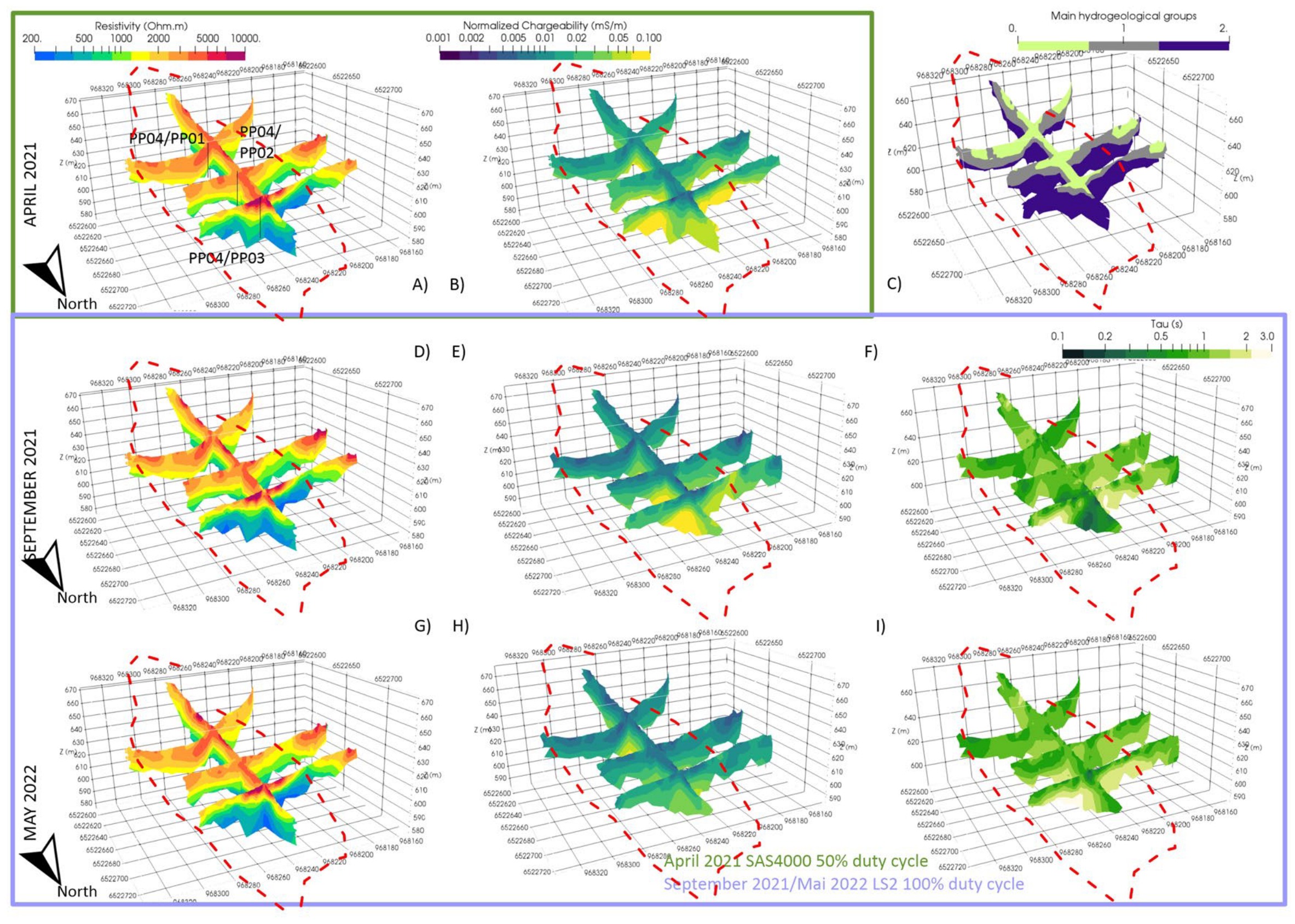}
\caption{Resistivity (A, D, G), normalized chargeability (B, E, H) and mean relaxation time (F, I) models obtained from the time-lapse inversion for the April 2021, September 2021 and May 2022 campaigns. (C) Main hydrogeological units defined on the basis of inclinometer data, piezometric measurements and geophysical models. Black lines represent the location of the further extracted vertical profiles.}
\label{fig:models}
\end{figure}

\begin{figure}[htbp]
\centering
\includegraphics[width=\textwidth]{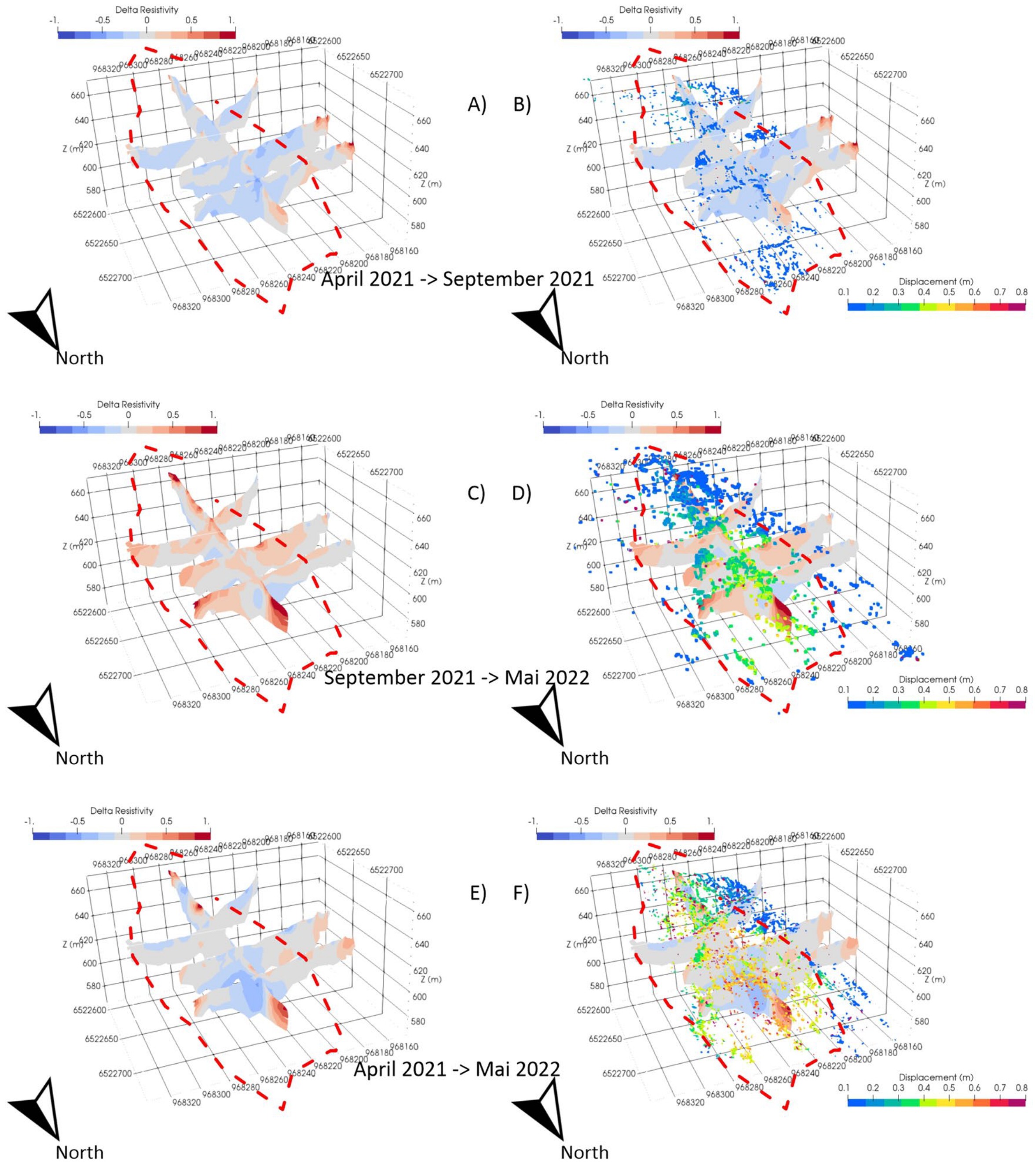}
\caption{Relative resistivity variations (A, C, E) and surface displacements obtained via TSM methodology (B, D, F). Relative variations are expressed in \% and displacements in meters.}
\label{fig:rhovar}
\end{figure}

\begin{figure}[htbp]
\centering
\includegraphics[width=\textwidth]{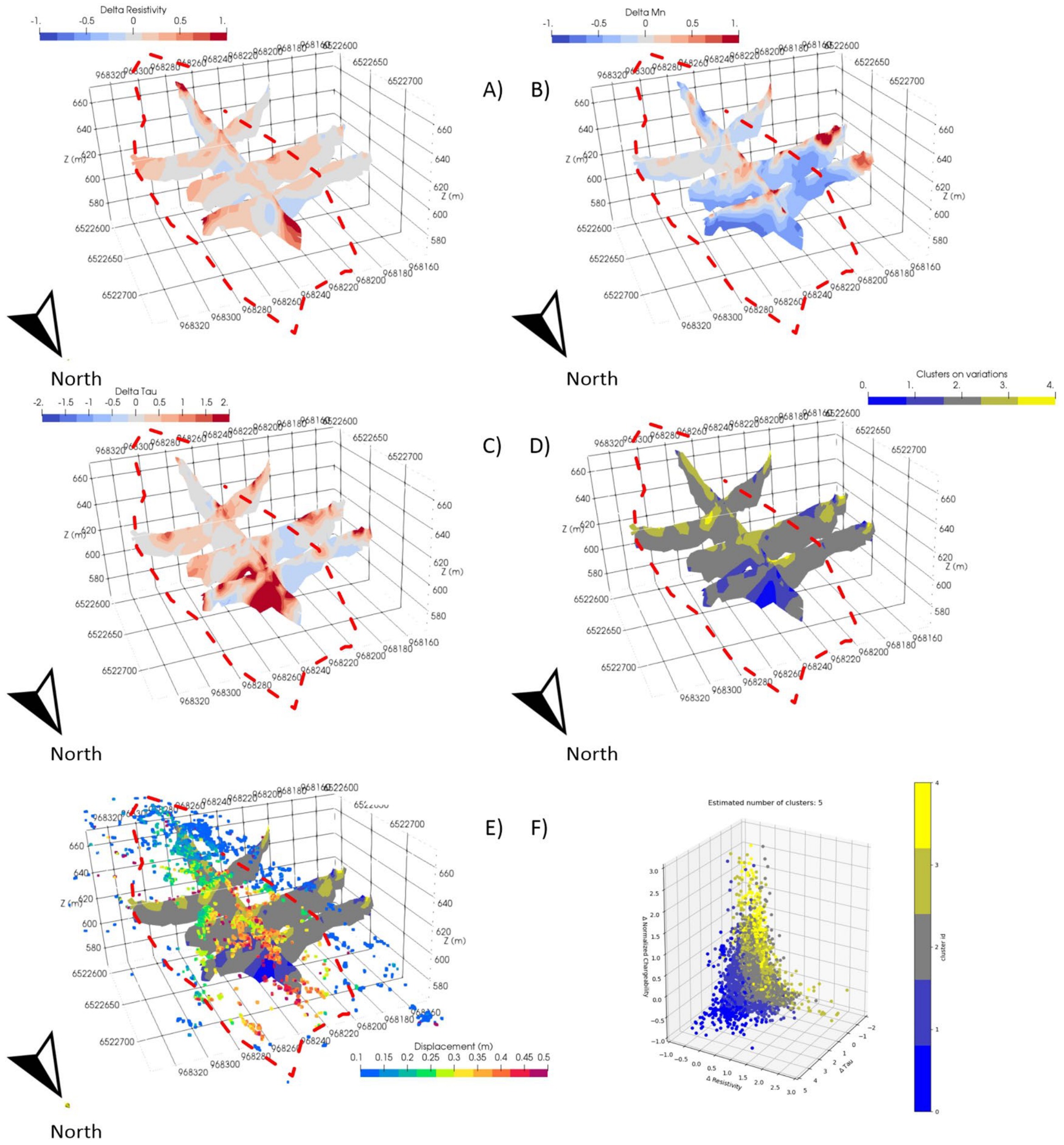}
\caption{Relative resistivity (A), normalized chargeability (B), mean relaxation time (C) variations between September 2021 and May 2022. The results of the clustering applied on relative variations and its confrontation to surface displacements is presented on panels (D), (E), and (F). The relative variations are defined as $\relvar$, where September 2021 is used as reference.}
\label{fig:var}
\end{figure}

\begin{figure}[htbp]
\centering
\includegraphics[height=0.6\textheight]{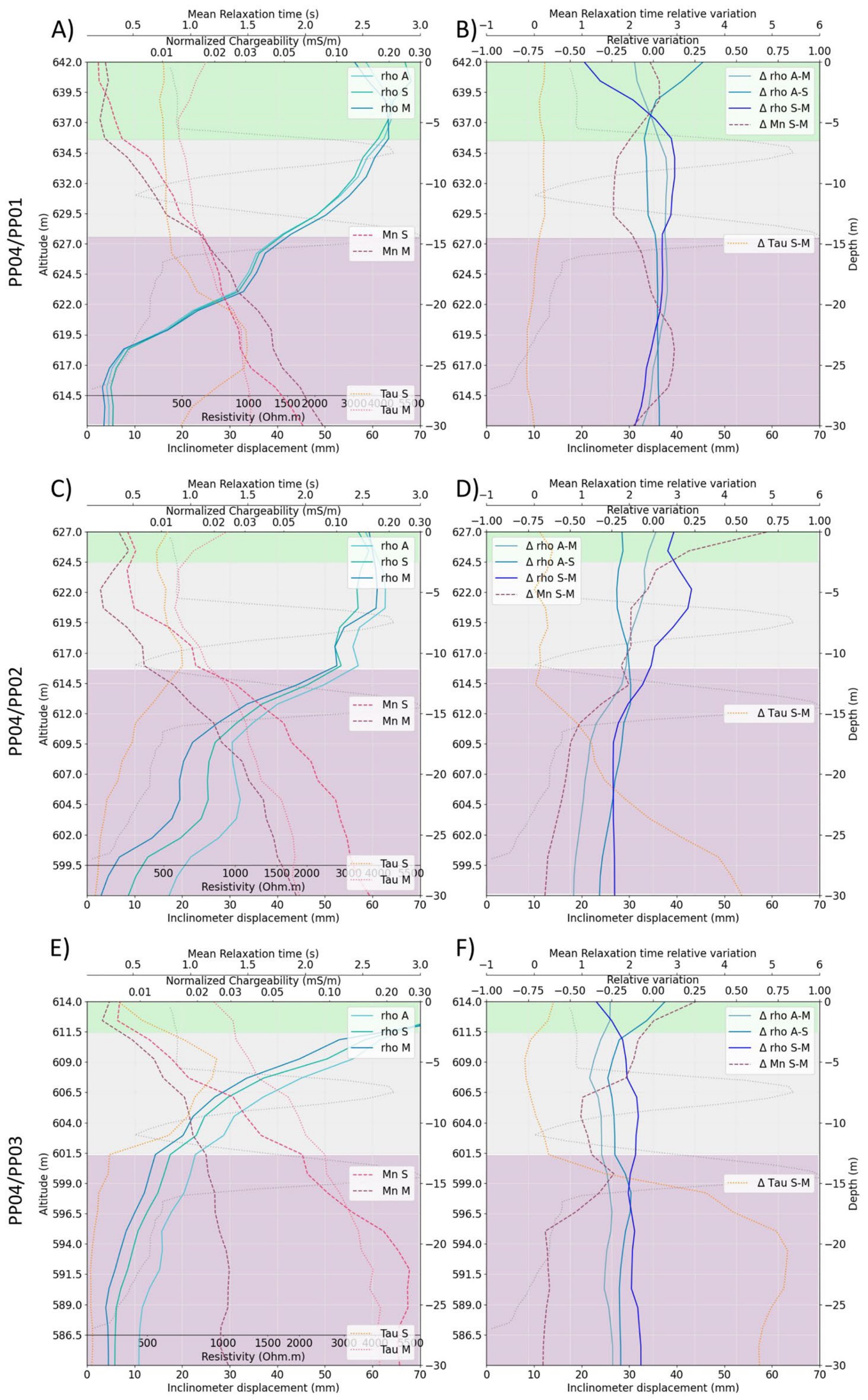}
\caption{Resistivity (rho in Ohm.m), normalized chargeability (Mn in mS/m) and mean relaxation times (Tau in s) at the three intersections between PP04 and PP01 (A), PP02 (C), and PP03 (E), respectively, from top to bottom of the landslide (location on Fig.~\ref{fig:models}). Relative resistivity, normalized chargeability and mean relaxation time variations at the three intersections between PP04 and PP01 (B), PP02 (D) and PP03 (F), respectively, from top to bottom of the landslide (relative variations are defined as $\relvar$ and have therefore no units). Please note that relative variations scale is different for B panel compared to D and F to ensure readability. The displacements measured at the inclinometer in mm are shown in grey in the background. The letters A, S and M in the captions stand for April 2021, September 2021 and May 2022. The green, grey, and purple backgrounds correspond to the three main hydrogeological units defined from the interpretation of the inclinometer, piezometric and geophysical data.}
\label{fig:profiles}
\end{figure}

Time-lapse ERT and IP inversion exhibits satisfying results; RMS ranges from 4.6 to 8.5\% and 9.6 to 12.3\% for ERT and IP inversion, respectively. The resulting cross-plots are shown on Fig.~\ref{fig:xplot}. The residuals for IP inversion are more spread than for ERT inversion which is consistent with the fact that IP data are more affected by noise than ERT data. For the DD, the mean and median values of the $\chi^2$ are 12 and 5 for the September 2021 data set and 5 and 3 for the May 2022 data set, respectively, which is satisfying. In the framework of the MGL, we do not expect any metallic particles within the materials, and we consider that the normalized chargeability is a proxy for surface conductivity and gives information about the amount of fine particles (Mao et al. 2016). The resistivity models exhibit high resistivities in the first 5~m depth ($>$4000~Ohm.m). Resistivity is gradually decreasing until 30~m depth to reach values between 200 and 500~Ohm.m. Low values are reached quicker in the downstream part of the MGL (Fig.~\ref{fig:models}A, D, G). Normalized chargeabilities values range between 0.001 and 0.3~mS/m. Normalized chargeabilities are less than 0.01~mS/m in the first 5~m depth and then increase with depth to reach 0.3~mS/m in the downstream part of the landslide in May 2022 (Fig.~\ref{fig:models}B, E, H). The relaxation time constant model exhibits values in the order of 1~s in the first 15~m depth. The values tend to decrease with depth in September 2021 and increase with depth in May 2022 in the downstream part of the landslide (Fig.~\ref{fig:models}F, I). We extracted 3 main hydrogeological units (HU) on the basis of inclinometer data, piezometric measurements, resistivity and normalized chargeability models (Fig.~\ref{fig:models}C). The first one (HU 0, green) consists in the first 5~m depth of the landslide made up of very loose material; it exhibits very high resistivities, low chargeabilities and variable relaxation time constants. Between 5 and 15~m depth, a second body (HU 1, grey) is defined where resistivities and chargeabilities are intermediate. This agrees with quite remobilized materials, moderately altered micaschists. At depth more than 15~m, materials with low resistivity and high chargeability are observed and constitute the third body (HU 2, purple). The latter suggests material with important surface conductivities and agrees with less damaged micaschists. The resistivity variations are represented along the corresponding surface displacements obtained via the TSM method over the considered period of time (Fig.~\ref{fig:rhovar}). The variations are defined as $\relvar$. The normalized chargeability and mean relaxation time variations are only considered between September 2021 and May 2022 when acquisition parameters and acquisition material are comparable (Figs.~\ref{fig:var} and \ref{fig:profiles}). The time-lapse inversion observations are the following:
\begin{itemize}
\item Resistivity tends to globally decrease within the landslide between April and September 2021. The surface displacements associated remain less than 0.1~m (Fig.~\ref{fig:rhovar}).
\item Between September 2021 and May 2022, the resistivity tends to globally increase in the first 15~m depth and decrease beyond 15~m depth in the downstream part of the landslide. The surface displacements associated are in the order of 0.3--0.4~m in the center/center-east-downstream part of the landslide. These results are in agreement with field observations suggesting that the landslide is more active in winter and spring (Fig.~\ref{fig:rhovar}). The normalized chargeability generally decreases between September and May for depths more than 15~m (Fig.~\ref{fig:var}), the decrease being more important in the center/center-east-downstream part of the landslide (up to 80\%) than in the uphill part (less than 20\% and more often less than 10\%). In the subsurface, it generally increases or stays rather stable in the upstream part of the landslide. The mean relaxation time is relatively stable in the uphill part of the landslide. In the downstream part, it slightly decreases at the outer rings, but it exhibits a significant increase of more than 200\% in the center/center-east-downstream part of the landslide (Figs.~\ref{fig:var} and \ref{fig:profiles}).
\item Between April 2021 and May 2022, resistivity has increased in the center/center east-downstream part of the landslide. The surface displacements associated are the order of 0.3--0.4~m in the eastern part of the landslide and can reach 0.8~m in the center/center-east-downstream part of the landslide. Resistivity variations are globally less than 20\% in absolute value (Fig.~\ref{fig:rhovar}).
\end{itemize}

In the HU 1, the resistivity and mean relaxation time tend to increase (from 0 to 20\% and 0 to 50\%, respectively) and the normalized chargeability tends to decrease (10 to 50\%) between September 2021 and May 2022. Yearly resistivity variations remain quasi-null. The variations observed in the HU 0 are more chaotic. So, in the first 15~m of the landslide, the variations are season-sensitive and variations from 1 year to the other remain quite small. The HU 2 exhibits few variations in the upstream part of the landslide (less than 10\%), whereas a resistivity and normalized chargeability decrease (up to 25 and 50\% respectively) and a strong mean relaxation time increase (more than 200\%) between September 2021 and May 2022 are observed in the downstream part (Figs.~\ref{fig:var} and \ref{fig:profiles}). Between the end of the summer and the spring, the piezometer level usually changes from about 30 to 15~m, meaning that this deeper geological layer of the landslide fills up with water during the winter and the spring and drains during the summer. In this hydrogeological unit, we observe seasonal variations but also a non-negligible permanent resistivity variation in the center/center-east-downstream part. This suggests that some water could be stored for a longer time at the downstream of the landslide. To help the reading of the variations results, a GMM clustering is applied to the variations of resistivity, normalized chargeability and mean relaxation time (Fig.~\ref{fig:var}D--F). Five groups are extracted, the groups 0 and 1 correspond to groups where the resistivity and normalized chargeability have decreased, and the mean relaxation time has significantly increased between September 2021 and May 2022. They are situated directly above the surface area where surface displacements are the strongest between September 2021 and May 2022. The group 2 exhibits low variations for the 3 variables, and the groups 3 and 4 a resistivity decrease, a normalized chargeability increase and a decrease of the mean relaxation time. We suggest that the groups 3 and 4 correspond to highly altered material very sensitive to seasonal variations and extreme events (fast snow melting, heavy rains) and represent 17,500~m$^3$. The groups 0 and 1 correspond to materials subject to seasonal and permanent variations, the latter controlling the overall large scale dynamics of the landslide. Groups 0 and 1 correspond to a volume of 6000 and 11,000~m$^3$, respectively.

\section{Discussion}

\subsection{IP methodology}

The measured TDIP data appear to be of good quality and consistent from one campaign to the other. The resulting normalized chargeabilities and mean relaxation times and their variations are spatially consistent which suggests that the results are trustworthy. However, we recorded field decay curves only on a limited time range (0.05 to 8~s). It implies a known information loss at early times (i.e., highest frequencies) and later times (i.e., lowest frequencies). Even if some useful information did come out from the time range used here, we suggest that more information about the spectral content of the IP response could be extracted with the use of a more elaborated post-processing technique as the one described by Olsson et al. (2016). This work has not been performed here and could be the subject of the further development of a freely available module. It has been revealed that the pulse length influences the magnitude and the shape of the decay curves (Mao et al. 2016; Olsson et al. 2019), so we choose to work with a fixed pulse length. We did not study the influence of changing the pulse length on our data mainly because of survey duration issues. Considering these limitations, we interpreted the RTD and extracted information (mean relaxation time, $\tau$) carefully with full knowledge and only qualitatively.

\subsection{Relation between variables and salinity and water content}

The interpretation of the DC and IP variables variations through the seasons implies to understand how each variable evolves with the water and salinity contents.

\begin{figure}[htbp]
\centering
\includegraphics[width=0.8\textwidth]{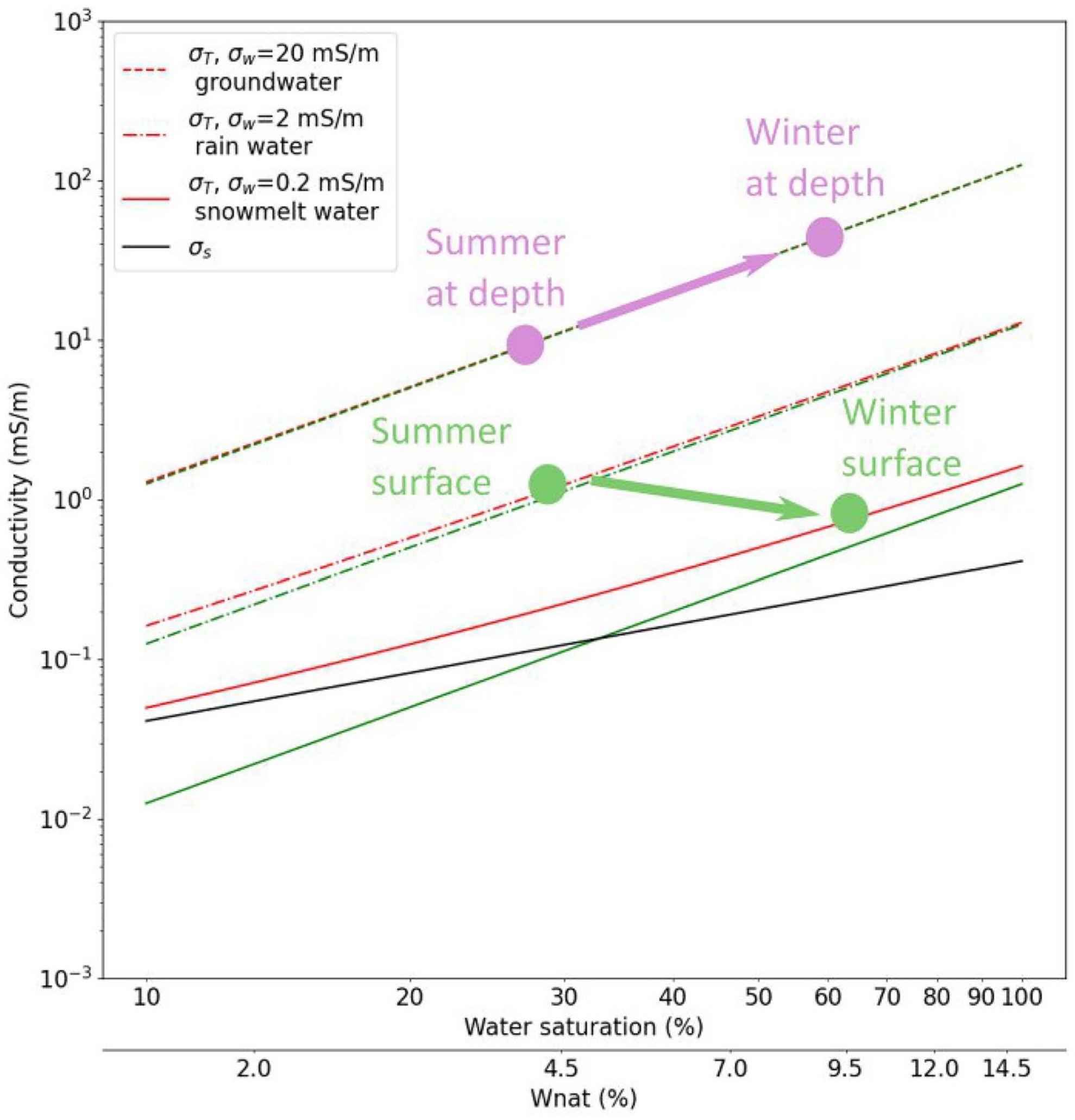}
\caption{Theoretical influence of water saturation on total, hydraulic and surface conductivity for different water conductivities for a typical material made of silt or clay silt. Red lines represent the total (bulk) DC conductivity ($\sigma_T$), green lines represent the electrolytic conductivity ($\sigma_H$), and black line represents surface conductivity ($\sigma_s$). $\sigma_w$ stands for water conductivity. Green and purple arrows represent the possible evolution at the surface and at 20~m depth within the MGL from the end of the summer to the spring. We consider the total DC conductivity as $\sigma_T = \sigma_H + \sigma_s = \theta^m \sigma_w + \theta^{m-1} \rho_g (B - \lambda) CEC$, where $\theta$ is the volumetric water content, $m$ the saturation exponent further set to 2, $\rho_g$ the grain density (kg.m$^{-3}$), $B$ and $\lambda$ the apparent mobility of counterions for surface conduction and polarization, respectively, and set to $3.1 \times 10^{-9}$ and $3 \times 10^{-10}$~m$^2$\,s$^{-1}$\,V$^{-1}$ and CEC the cation exchange capacity in meq/100g further set to 20 (Revil et al. 2020).}
\label{fig:theory}
\end{figure}

The nature and origin of seasonal variations of resistivity within landslides are strongly site and geology dependent (Supper et al. 2014; Gance et al. 2016; Palis et al. 2017a, b; Uhlemann et al. 2017; Holmes et al. 2020). What is more, the resistivity variations observed in clayey landslides are quite tenuous, generally less than 10\% (Gance et al. 2016). Between September and May at the MGL, we observed either a resistivity increase or decrease whether we consider surface or deep landslide material, respectively. We assume from the field observations and the piezometric levels that the water saturation increases from summer to winter within the landslide even if we cannot quantify it precisely. To illustrate and better understand the resistivity variations through the seasons in the different parts of the MGL, we computed the theoretical evolution of the total conductivity as a function of water saturation at different water conductivities (i.e., different salinities) for a typical clay/silt material taking the surface conductivity into account following the Revil et al. (2020) model. Figure~\ref{fig:theory} presents the evolution of the bulk conductivity with saturation for pores filled of snow melt water (0.2~mS/m), rain water (2~mS/m) or ground water (about 20~mS/m with almost no seasonal variations from publicly available water quality measurements for the different sources in the Arly valley). When water saturation increases within the landslide during winter and spring, the conductivity increases in the deeper parts of the landslide due to an increase of water content of constant medium salinity (purple arrow on the Fig.~\ref{fig:theory}). Meanwhile, it decreases within the subsurface material due to the very low conductivity of the brought fluids (green arrow). In this latter case, we also note that the surface conductivity is not negligible, so the Archie's law for water saturation computation is not applicable.

The relation between normalized chargeability or relaxation time and saturation is strongly material dependent, so the polarization models inferred to explain these relations are not universal (Titov et al. 2004; Kemna et al. 2005; Binley et al. 2005; Ghorbani et al. 2009; Revil 2012; Breede et al. 2012; Okay et al. 2013; Abdulsamad et al. 2017; Zhang et al. 2019; Martin et al. 2022). In our case, the total chargeability of our mixture of damaged/crushed micaschists and silts never exceeds 40~mV/V, and no metallic particles are observed. Our media might not act as pure clay neither like material mixed with metallic particles but can rather be compared to mixtures of sand and clay. The polarization mechanisms hence inferred could be a combination of the short narrow pore model (Titov et al. 2004) and a possible rearrangement of clay particles in the pore space (Breede et al. 2012). In this type of medium, the relaxation time constant increases with saturation and the normalized chargeability increases and then decreases after reaching a certain saturation threshold. The mean relaxation time is also linked to the characteristic diffusion length related to the pore size or grain size. Salinity also affects the polarization mechanisms (Revil 2012; Lévy et al. 2019b; Martin et al. 2022). Its influence strongly depends on the dominant polarization and conduction mechanisms. Experiments realized on sands, sandstones and clay agreed on the fact that the polarization tends to increase with salinity until reaching a threshold and decreases at very high salinities (Revil and Skold 2011; Weller and Slater 2012; Weller et al. 2013; Hördt et al. 2016; Mendieta et al. 2021). The salinities measured on our field range in the very low salinity ranges of these experiments. Hence, we would expect the normalized chargeability to decrease when the salinity decreases.

Between September 2021 and May 2022, in the HU units 0 and 1, the resistivity and the mean relaxation time slightly increase, and the normalized chargeability increases in parallel to an important amount of rain and snowmelt water at the surface. Considering the previous bibliographic observations, this is consistent with the saturation increase of a sand/clay mixture from a very low starting saturation. As the normalized chargeability did not decrease with the salinity decrease (attested by the resistivity increase), we conclude that the salinity effect is not the dominant effect on polarization in these units. In the shallowest areas, the variations are heterogeneous showing a strong heterogeneity in clay content and/or salinity distribution. In the HU 2, the resistivity and the normalized chargeability decrease and the mean relaxation time constant strongly increases. The latter suggests a significant saturation increase if we assume a constant or almost constant pore size distribution through time, confirmed by the piezometric level observed. The decrease of the normalized chargeability could then be explained by a saturation increase from an already partially saturated state. Indeed, the resistivity decrease suggests that the salinity did not decrease significantly and is not responsible for the normalized chargeability decrease. Considering all the previous observations, we conclude that the main sources of feeding water in the subsurface units (HUs 0 and 1) are the rain and snow melt events, whereas in the HU 2 (deeper unit), the water circulating within the massif is the main source of additional fluids.

\subsection{Permeability estimation}

\begin{figure}[htbp]
\centering
\includegraphics[height=0.75\textheight]{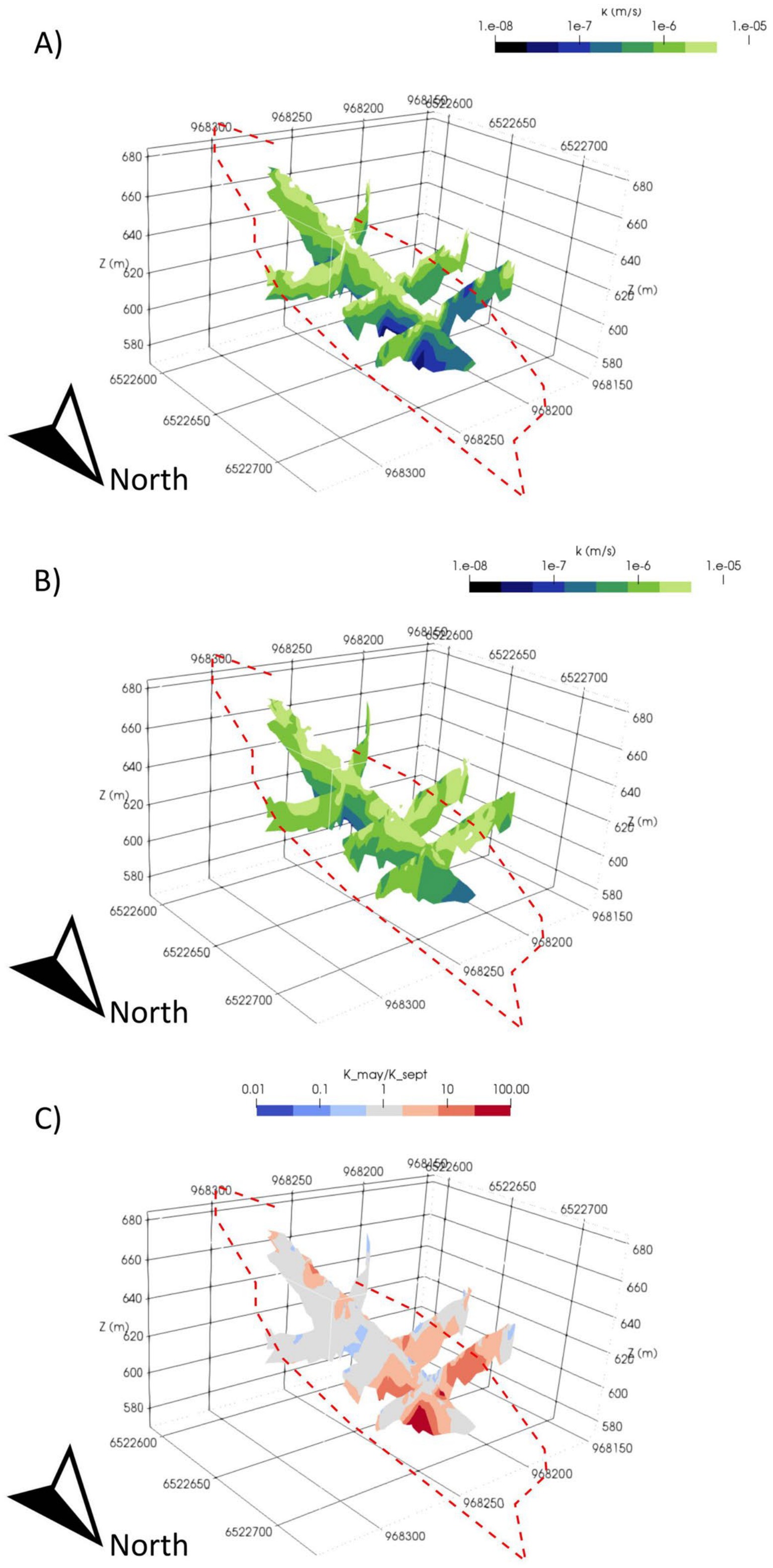}
\caption{Permeability models for September 2021 (A), May 2022 (B) and their ratio (C) extracted from Weller et al. (2015) relationship.}
\label{fig:perm}
\end{figure}

\begin{figure}[htbp]
\centering
\includegraphics[width=\textwidth]{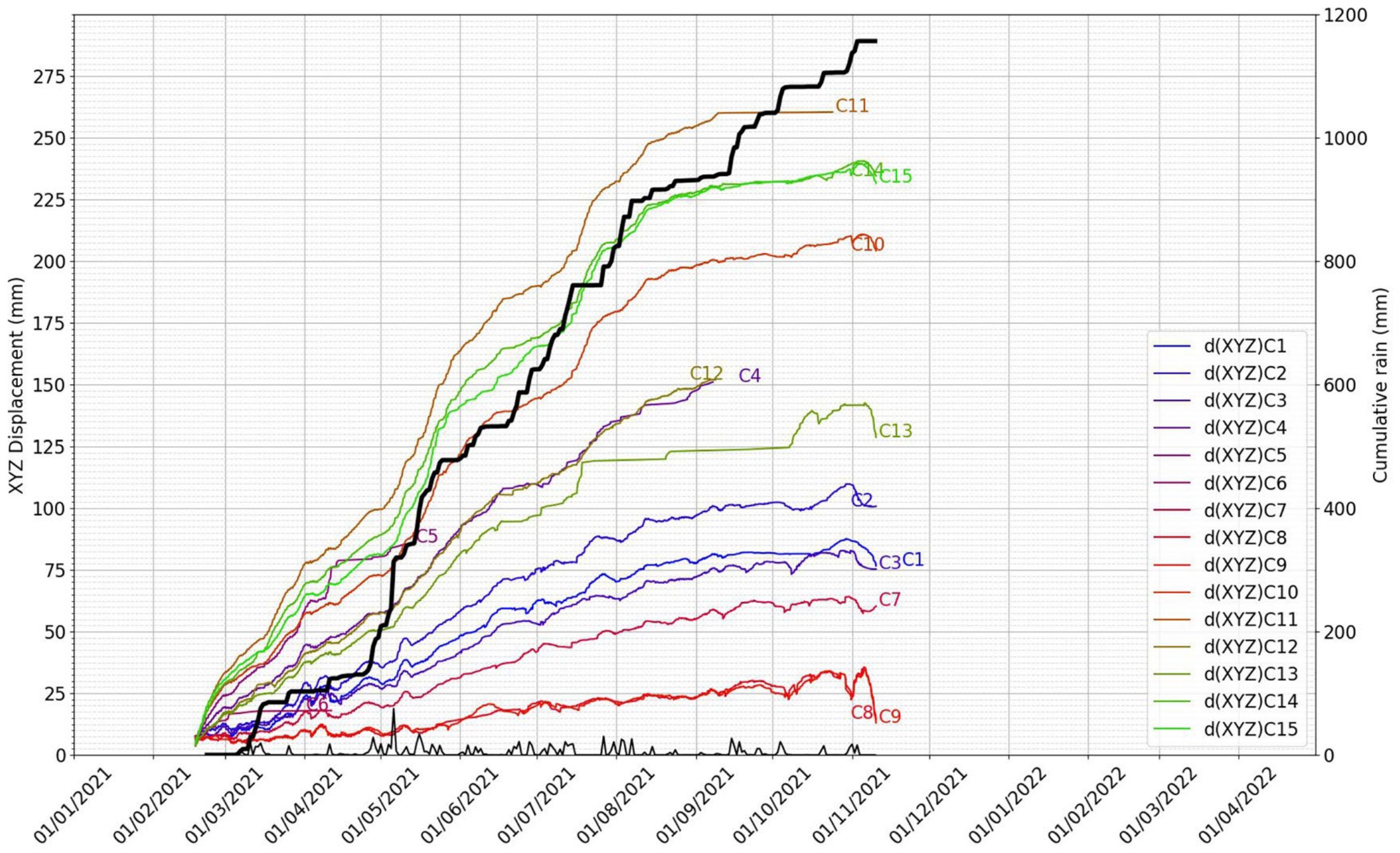}
\caption{Surface displacements measured from February to November 2021 using an automated theodolite. Black curves represent rain in mm/day (thin black line) and cumulated rain (bold black line) in mm.}
\label{fig:disp}
\end{figure}

IP measurements are related to the surface area of the pore network, consequently to the pore size, and help to predict the permeability (Börner et al. 1996; Slater 2007; Revil 2012; Weller et al. 2015). Recently, several studies proposed the use of laboratory calibrated relationships between permeability and IP parameters to estimate the permeability at the field scale (Fiandaca et al. 2018; Maurya et al. 2018; Revil et al. 2020). In all the studies, the estimation of the permeability implies to consider some fixed parameters (like porosity or water conductivity) and/or some assumption about petrophysical models. Some studies also proposed to link the permeability with the relaxation time, but the relationships are strongly dependent on the geology (Binley et al. 2005; Tong et al. 2006b; Zisser et al. 2010). Considering these limitations, we decided to apply the relationship proposed by Weller et al. (2015) and applied on the field by Orozco et al. (2022) to infer an order of magnitude of the permeability and its evolution through space and time using the normalized chargeability as a proxy of the imaginary conductivity. It avoids the use of the formation factor which cannot be properly spatially defined in our case. Hydraulic conductivity is then defined as:
\begin{equation}
K = \frac{g.d_w}{\mu}\,\frac{3.47 \times 10^{-16}\,\sigma_{DC}^{1.11}}{\sigma''^{2.41}}
\label{eq:k}
\end{equation}
where $g$ is the gravitational constant, and $d_w$ and $\mu$ are groundwater density and dynamic viscosity in kg.m$^{-3}$ and kg.m$^{-1}$\,s$^{-1}$, respectively. The relationship between the normalized chargeability and the quadrature conductivity is related to the number of decades of the frequency range considered such as $Mn \approx \alpha\sigma''$ with $\alpha \approx \frac{2}{\pi}\ln(A)$, where $A = 10^k$, $k$ being the number of decades between minimum and maximum acquisition frequencies. In our case, $\alpha \approx 4$. Hydraulic conductivity decreases from $10^{-5}$~m/s in the subsurface of the MGL to $10^{-8}$~m/s at 30~m depth within the landslide (Fig.~\ref{fig:perm}). After heavy rains or quick snow melt episodes, resurgences appear within a few days at the base of the landslide, suggesting that the loose surface material is permeable. Considering a 5~m depth layer, the vertical hydraulic conductivity would be the order of $10^{-5}$~m/s. The saturation and the resulting pore pressure increase of these loose subsurface materials generate rock flows almost permanently during the rainy season. Figure~\ref{fig:disp} represents the rain and cumulative rain measured at the site along with the surface displacements during the year 2021. The cross-correlation between the signals exhibits a maximum at a lag of 25 to 30 days for the targets 4 to 15 situated in the active part of the landslide. Considering a 15~m depth sliding surface, the resulting vertical hydraulic conductivity would be the order of $10^{-7}$~m/s. Here, we underline the fact that the orders of magnitude of hydraulic conductivity estimated from the field observations are consistent with the estimation made from the geophysical data.

\subsection{Hydrogeological model for the MGL}

\begin{figure}[htbp]
\centering
\includegraphics[height=0.72\textheight]{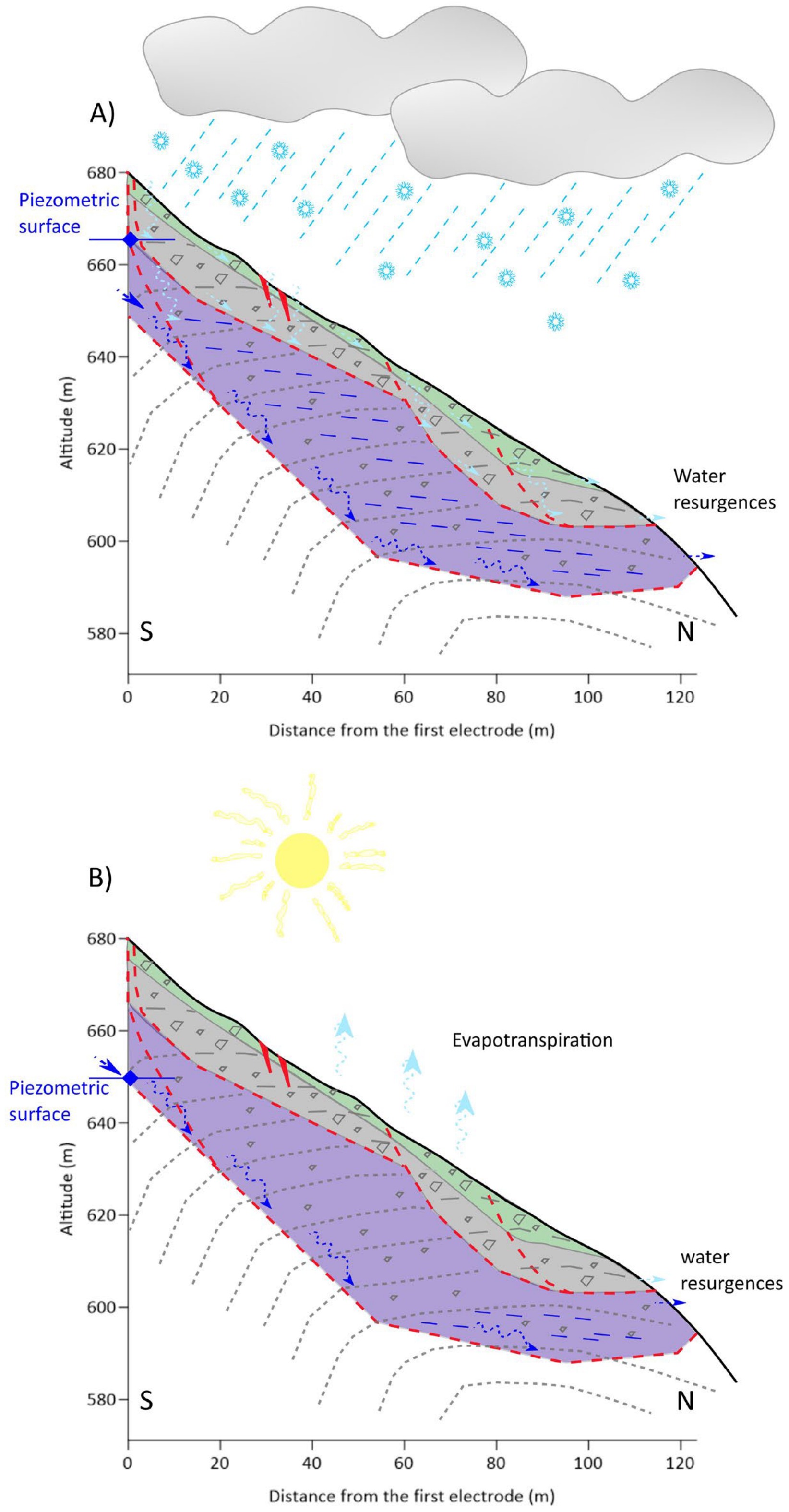}
\caption{Simplified cross-section of the MGL landslide along the PP04 profile showing subsurface hydrogeological fluid dynamics through the seasons (A December to May and B June to November). The three main bodies (green, grey, purple) are the three main hydrogeological bodies identified in Section~\ref{sec:results}. The damage degree of the micaschists is increasing with depth. Toppling is beginning at the interface between the bedrock and the purple body, and damage progressively increases toward the surface in the two other units. Red dashed lines represent the landslide surfaces inferred from inclinometer data and geophysical data. The purple body acts as a semi-permeable aquifer unit, and piezometric surface varies within this body. Soil moisture accumulation is possible at the bottom of the landslide. Red triangles represent open cracks acting as permeable barriers.}
\label{fig:model}
\end{figure}

The MGL results from the toppling failure of the sub-vertical micaschists. The latter are strongly foliated rocks composed of sheets of phyllosilicates. In such materials, hydraulic conductivity is strongly dependent on the direction and so is polarization (Kelly and Reiter 1984; Cosenza et al. 2007; Börner et al. 2018; Abdulsamad et al. 2019). In the case of the MGL, the horizontal bending of the micaschists generated by the toppling failure would generate low vertical hydraulic conductivities in the 15 to 30~m depth unit (HU 2). But this same layer could be more permeable in the opposite direction which would allow the fluids from within the massif to circulate. HU 2 is acting as an aquifer and undergoes seasonal water table variations (Fig.~\ref{fig:model}). During winter and spring, water infiltrates from the upper massif and gathers in HU 2 at the downstream of the landslide. As a consequence, the pore pressure increases and the available space for electrical charges conduction widens, which lead to a decrease of normalized chargeability and resistivity, an increase in mean relaxation time and an increase in permeability. At this point, general mass movements of the landslide appear. When HU 2 is pressurized, the units above are more prone to destabilization and respond quickly to rain and snow melt events. We suggest that the observed surface displacements result from the combination of rain/snow melt infiltration and underground water accumulation at depth.

\section{Conclusion}

This paper presents the results of the first DC and TDIP time-lapse inversion performed on an active landslide. A methodology developed to get the spectral information from IP data through a Debye decomposition has been applied in order to extract as much information as possible from the TDIP data. The resistivity variations are quite small within the landslide and do not yield significant information about the unstable volumes. In contrast, the IP data allowed us to delineate the geometry and volume of the MGL unstable slope. The results of the time-lapse ERT and TDIP are very well correlated with the surface displacements obtained via the TSM method. The use of a non-supervised classification method enables to build groups on the basis of the temporal evolution of the variables without formulating any hypothesis about the physical mechanisms or using petrophysical models. The time-lapse inversion of the geophysical data and the clustering allows us to discriminate between a weather-sensitive subsurface volume with potentially fast reaction times to extreme meteorological events and a deeper volume affected by seasonal water level variation but also exhibiting permanent variations. The first one represents at least 17,500~m$^3$ of materials, and the latter, situated above the area where surface displacements are the most important, represents a total of 17,000~m$^3$ potentially unstable materials. These results are critical for the design of security systems to monitor the MGL. TDIP time-lapse data have the potential to provide additional information about the water content and the geometry of fluid flow circulations within landslides as long as we understand how the considered variables evolve with water content and/or salinity. Hence, TDIP appears to be a very promising tool in landslide characterization and dynamics assessment. Note that the possibilities could be enhanced by the improvement of the data processing methodology and the use of complementary geophysical data such as refraction seismic data.

This paper did not focus on the theory of the IP, neither on the establishment of a petrophysical model for the studied site. In order to better understand the physical processes at stake, especially the polarization mechanisms at work from a polarization theory point of view, more laboratory data and study would be needed.

\section*{Acknowledgements}

This work has been funded by the POIA-FEDER European program in the framework of the MIROIR project (PA0020589). We would like to thank the BRGM and especially Edouard Equilbey for supporting the MIROIR project. We are also grateful to the CD73 and especially Anne Lescurier for providing data and feedbacks about the observed dynamics of the landslide. We would like to thank Nicolas Moreau, Charles Ribeyre and Mathilde Masson for their precious help on the field. We also warmly thank Thomas Günther and Tina Martin for their very useful advice. We also thank the two anonymous reviewers for their constructive comments which greatly helped to improve the quality of the manuscript.

\section*{Data availability}

The data are available upon request.

\section*{Declarations}

\textbf{Consent for publication.} The authors have the consent for publication of the data.

\noindent\textbf{Conflict of interest.} The authors declare no competing interests.

\section*{References}
\begingroup
\small
\setstretch{1.1}
\setlength{\parindent}{-1.5em}
\setlength{\leftskip}{1.5em}
\setlength{\parskip}{0.3em}
\sloppy

Abdulsamad F, Florsch N, Camerlynck C (2017) Spectral induced polarization in a sandy medium containing semiconductor materials: experimental results and numerical modelling of the polarization mechanism. Near Surface Geophysics 15(6):669--683

Abdulsamad F, Revil A, Ghorbani A et al (2019) Complex conductivity of graphitic schists and sandstones. J Geophys Res Solid Earth 124(8):8223--8249

Binley A, Kemna A (2005) DC resistivity and induced polarization methods. In: Hydrogeophysics. Springer, p 129--156

Binley A, Slater LD, Fukes M et al (2005) Relationship between spectral induced polarization and hydraulic properties of saturated and unsaturated sandstone. Water Resourc Res 41(12)

Börner F, Schopper J, Weller A (1996) Evaluation of transport and storage properties in the soil and groundwater zone from induced polarization measurements. Geophys Prospect 44(4):583--601

Börner JH, Girault F, Bhattarai M et al (2018) Anomalous complex electrical conductivity of a graphitic black schist from the Himalayas of central Nepal. Geophys Res Lett 45(9):3984--3993

Breede K, Kemna A, Esser O et al (2012) Spectral induced polarization measurements on variably saturated sand-clay mixtures. Near Surface Geophysics 10(6):479--489

Brunet P, Clément R, Bouvier C (2010) Monitoring soil water content and deficit using electrical resistivity tomography (ERT) -- a case study in the Cevennes area, France. J Hydrol 380(1--2):146--153

Carrier A, Bottelin P, Fabre L et al (2022) Damage assessment of supporting pillars in an underground cave using joint inversion of electrical resistivity and P-wave velocity, Burgundy (France). Pure Appl Geophys 179(1):45--67

Cascini L, Scoppettuolo MR, Babilio E (2022) Forecasting the landslide evolution: from theory to practice. Landslides 1--13

Chambers J, Gunn D, Wilkinson P et al (2014) 4D electrical resistivity tomography monitoring of soil moisture dynamics in an operational railway embankment. Near Surface Geophysics 12(1):61--72

Coperey A, Revil A, Abdulsamad F et al (2019) Low-frequency induced polarization of porous media undergoing freezing: preliminary observations and modeling. J Geophys Res Solid Earth 124(5):4523--4544

Cosenza P, Ghorbani A, Florsch N et al (2007) Effects of drying on the low-frequency electrical properties of Tournemire argillites. Pure Appl Geophys 164(10):2043--2066

Dahlin T, Leroux V, Nissen J (2002) Measuring techniques in induced polarisation imaging. J Appl Geophys 50(3):279--298

Delforge D, Watlet A, Kaufmann O et al (2021) Time-series clustering approaches for subsurface zonation and hydrofacies detection using a real time-lapse electrical resistivity dataset. J Appl Geophys 184:104203

Desrues M, Malet JP, Brenguier O et al (2019) TSM-tracing surface motion: a generic toolbox for analyzing ground-based image time series of slope deformation. Remote Sensing 11(19):2189

Desrues M, Malet JP, Brenguier O et al (2022) Landslide kinematics inferred from in situ measurements: the Cliets rock-slide (Savoie, French Alps). Landslides 19(1):19--34

Doetsch J, Ingeman-Nielsen T, Christiansen AV et al (2015) Direct current (DC) resistivity and induced polarization (IP) monitoring of active layer dynamics at high temporal resolution. Cold Reg Sci Technol 119:16--28

Eberhardt E, Thuro K, Luginbuehl M (2005) Slope instability mechanisms in dipping interbedded conglomerates and weathered marls -- the 1999 Rufi landslide, Switzerland. Eng Geol 77(1--2):35--56

Fiandaca G, Ramm J, Binley A et al (2013) Resolving spectral information from time domain induced polarization data through 2-D inversion. Geophys J Int 192(2):631--646

Fiandaca G, Madsen LM, Maurya PK (2018) Re-parameterisations of the Cole-Cole model for improved spectral inversion of induced polarization data. Near Surface Geophysics 16(4):385--399

Gance J, Malet JP, Dewez T et al (2014) Target detection and tracking of moving objects for characterizing landslide displacements from time-lapse terrestrial optical images. Eng Geol 172:26--40

Gance J, Malet JP, Supper R et al (2016) Permanent electrical resistivity measurements for monitoring water circulation in clayey landslides. J Appl Geophys 126:98--115

Gazoty A, Fiandaca G, Pedersen J et al (2012) Mapping of landfills using time-domain spectral induced polarization data: the Eskelund case study. Near Surface Geophysics 10(6):575--586

Ghorbani A, Cosenza P, Revil A et al (2009) Non-invasive monitoring of water content and textural changes in clay-rocks using spectral induced polarization: a laboratory investigation. Appl Clay Sci 43(3--4):493--502

Günther T, Rücker C, Spitzer K (2006) Three-dimensional modelling and inversion of dc resistivity data incorporating topography -- II. Inversion. Geophys J Int 166(2):506--517

Gurin G, Tarasov A, Ilyin Y et al (2013) Time domain spectral induced polarization of disseminated electronic conductors: laboratory data analysis through the Debye decomposition approach. J Appl Geophys 98:44--53

Guzzetti F, Peruccacci S, Rossi M et al (2008) The rainfall intensity-duration control of shallow landslides and debris flows: an update. Landslides 5(1):3--17

Hayley K, Bentley LR, Gharibi M et al (2007) Low temperature dependence of electrical resistivity: implications for near surface geophysical monitoring. Geophys Res Lett 34(18)

Hellman K, Ronczka M, Günther T et al (2017) Structurally coupled inversion of ERT and refraction seismic data combined with cluster-based model integration. J Appl Geophys 143:169--181

Hibert C, Grandjean G, Bitri A et al (2012) Characterizing landslides through geophysical data fusion: example of the La Valette landslide (France). Eng Geol 128:23--29

Holmes J, Chambers J, Meldrum P et al (2020) Four-dimensional electrical resistivity tomography for continuous, near-real-time monitoring of a landslide affecting transport infrastructure in British Columbia, Canada. Near Surface Geophysics 18(Geoelectrical Monitoring):337--351

Hördt A, Bairlein K, Bielefeld A et al (2016) The dependence of induced polarization on fluid salinity and pH, studied with an extended model of membrane polarization. J Appl Geophys 135:408--417

Kelly W, Reiter P (1984) Influence of anisotropy on relations between electrical and hydraulic properties of aquifers. J Hydrol 74(3--4):311--321

Kemna A, Münch HM, Titov K et al (2005) Relation of SIP relaxation time of sands to salinity, grain size and hydraulic conductivity. In: Near Surface 2005 -- 11th European Meeting of Environmental and Engineering Geophysics, European Association of Geoscientists \& Engineers, pp cp--13

Kemna A, Binley A, Cassiani G et al (2012) An overview of the spectral induced polarization method for near-surface applications. Near Surface Geophysics 10(6):453--468

Lacroix P, Bièvre G, Pathier E et al (2018) Use of sentinel-2 images for the detection of precursory motions before landslide failures. Remote Sens Environ 215:507--516

Lajaunie M, Gance J, Nevers P et al (2019) Structure of the Séchilienne unstable slope from large-scale three-dimensional electrical tomography using a resistivity distributed automated system (R-DAS). Geophys J Int 219(1):129--147

Lévy L, Maurya P, Byrdina S et al (2019) Electrical resistivity tomography and time-domain induced polarization field investigations of geothermal areas at Krafla, Iceland: comparison to borehole and laboratory frequency-domain electrical observations. Geophys J Int 218(3):1469--1489

Lévy L, Weller A, Gibert B (2019b) Influence of smectite and salinity on the imaginary and surface conductivity of volcanic rocks. Near Surface Geophysics 17(6-Recent Developments in Induced Polarization):653--673

Lu N, Godt JW, Wu DT (2010) A closed-form equation for effective stress in unsaturated soil. Water Resour Res 46(5)

Madsen LM, Fiandaca G, Auken E et al (2017) Time-domain induced polarization -- an analysis of Cole-Cole parameter resolution and correlation using Markov Chain Monte Carlo inversion. Geophys J Int 211(3):1341--1353

Malet J, Durand Y, Remaître A et al (2007) Assessing the influence of climate change on the activity of landslides in the Ubaye valley. In: Proceedings of the International Conference on Landslides and Climate Change -- Challenges and Solutions, edited by: McInnes R, Jakeways J, Fairbank H, and Mathie E, Taylor \& Francis, London, pp 195--205

Malet JP, Laigle D, Remaître A et al (2005) Triggering conditions and mobility of debris flows associated to complex earthflows. Geomorphology 66(1--4):215--235

Mao D, Revil A, Hinton J (2016) Induced polarization response of porous media with metallic particles -- Part 4: detection of metallic and nonmetallic targets in time-domain-induced polarization tomography. Geophysics 81(4):D359--D375

Marescot L, Monnet R, Chapellier D (2008) Resistivity and induced polarization surveys for slope instability studies in the Swiss Alps. Eng Geol 98(1--2):18--28

Martin T, Günther T, Orozco AF et al (2020) Evaluation of spectral induced polarization field measurements in time and frequency domain. J Appl Geophys 180:104141

Martin T, Titov K, Tarasov A et al (2021) Spectral induced polarization: frequency domain versus time domain laboratory data. Geophys J Int 225(3):1982--2000

Martin T, Weller A, Behling L (2022) Desaturation effects of pyrite-sand mixtures on induced polarization signals. Geophys J Int 228(1):275--290

Maurya P, Fiandaca G, Christiansen A et al (2018) Field-scale comparison of frequency- and time-domain spectral induced polarization. Geophys J Int 214(2):1441--1466

Mendieta A, Jougnot D, Leroy P et al (2021) Spectral induced polarization characterization of non-consolidated clays for varying salinities -- an experimental study. J Geophys Res Solid Earth 126(4): e2020JB021125

Nordsiek S, Weller A (2008) A new approach to fitting induced-polarization spectra. Geophysics 73(6):F235--F245

Okay G, Cosenza P, Ghorbani A et al (2013) Localization and characterization of cracks in clay-rocks using frequency and time-domain induced polarization. Geophys Prospect 61(1):134--152

Oldenburg DW, Li Y (1994) Inversion of induced polarization data. Geophysics 59(9):1327--1341

Olsson PI, Dahlin T, Fiandaca G et al (2015) Measuring time-domain spectral induced polarization in the on-time: decreasing acquisition time and increasing signal-to-noise ratio. J Appl Geophys 123:316--321

Olsson PI, Fiandaca G, Larsen JJ et al (2016) Doubling the spectrum of time-domain induced polarization by harmonic de-noising, drift correction, spike removal, tapered gating and data uncertainty estimation. Geophysical Supplements to the Monthly Notices of the Royal Astronomical Society 207(2):774--784

Olsson PI, Fiandaca G, Maurya PK et al (2019) Effect of current pulse duration in recovering quantitative induced polarization models from time-domain full-response and integral chargeability data. Geophys J Int 218(3):1739--1747

Orozco AF, Steiner M, Katona T et al (2022) Application of induced polarization imaging across different scales to understand surface and groundwater flow at the Hofermuehle landslide. CATENA 219:106612

Palis E, Lebourg T, Tric E et al (2017a) Long-term monitoring of a large deep-seated landslide (La Clapiere, South-East French Alps): initial study. Landslides 14(1):155--170

Palis E, Lebourg T, Vidal M et al (2017b) Multiyear time-lapse ERT to study short- and long-term landslide hydrological dynamics. Landslides 14(4):1333--1343

Pedregosa F, Varoquaux G, Gramfort A et al (2011) Scikit-learn: machine learning in python. J Mach Learn Res 12:2825--2830

Ponziani F, Pandolfo C, Stelluti M et al (2012) Assessment of rainfall thresholds and soil moisture modeling for operational hydrogeological risk prevention in the Umbria region (central Italy). Landslides 9(2):229--237

Provost F, Michéa D, Malet JP et al (2022) Terrain deformation measurements from optical satellite imagery: the MPIC-OPT processing services for geohazards monitoring. Remote Sens Environ 274:112949

Revil A (2012) Spectral induced polarization of shaly sands: influence of the electrical double layer. Water Resour Res 48(2)

Revil A, Skold M (2011) Salinity dependence of spectral induced polarization in sands and sandstones. Geophys J Int 187(2):813--824

Revil A, Coperey A, Shao Z et al (2017) Complex conductivity of soils. Water Resour Res 53(8):7121--7147

Revil A, Ahmed AS, Coperey A et al (2020) Induced polarization as a tool to characterize shallow landslides. J Hydrol 589:125369

Ronczka M, Hellman K, Günther T et al (2017) Electric resistivity and seismic refraction tomography: a challenging joint underwater survey at Äspö Hard Rock Laboratory. Solid Earth 8(3):671--682

Rücker C, Günther T, Spitzer K (2006) Three-dimensional modelling and inversion of dc resistivity data incorporating topography -- I. Modelling. Geophys J Int 166(2):495--505

Rücker C, Günther T, Wagner FM (2017) pyGIMLi: an open-source library for modelling and inversion in geophysics. Computers \& Geosciences 109:106--123

Scoppettuolo M, Cascini L, Babilio E (2020) Typical displacement behaviours of slope movements. Landslides 17(5):1105--1116

Seigel HO (1959) Mathematical formulation and type curves for induced polarization. Geophysics 24(3):547--565

Slater L (2007) Near surface electrical characterization of hydraulic conductivity: from petrophysical properties to aquifer geometries -- a review. Surv Geophys 28(2):169--197

Slater L, Glaser D (2003) Controls on induced polarization in sandy unconsolidated sediments and application to aquifer characterization. Geophysics 68(5):1547--1558

Slater LD, Lesmes D (2002) IP interpretation in environmental investigations. Geophysics 67(1):77--88

Stumpf A, Malet JP, Delacourt C (2017) Correlation of satellite image time-series for the detection and monitoring of slow-moving landslides. Remote Sens Environ 189:40--55

Supper R, Ottowitz D, Jochum B et al (2014) Geoelectrical monitoring: an innovative method to supplement landslide surveillance and early warning. Near Surface Geophysics 12(1):133--150

Tarasov A, Titov K (2007) Relaxation time distribution from time domain induced polarization measurements. Geophys J Int 170(1):31--43

Titov K, Kemna A, Tarasov A et al (2004) Induced polarization of unsaturated sands determined through time domain measurements. Vadose Zone Journal 3(4):1160--1168

Titov K, Tarasov A, Ilyin Y et al (2010) Relationships between induced polarization relaxation time and hydraulic properties of sandstone. Geophys J Int 180(3):1095--1106

Tong M, Li L, Wang W et al (2006a) Determining capillary-pressure curve, pore-size distribution, and permeability from induced polarization of shaley sand. Geophysics 71(3):N33--N40

Tong M, Li L, Wang W et al (2006b) A time-domain induced-polarization method for estimating permeability in a shaly sand reservoir. Geophys Prospect 54(5):623--631

Travelletti J, Delacourt C, Allemand P et al (2012) Correlation of multi-temporal ground-based optical images for landslide monitoring: application, potential and limitations. ISPRS J Photogramm Remote Sens 70:39--55

Tso CHM, Kuras O, Wilkinson PB et al (2017) Improved characterisation and modelling of measurement errors in electrical resistivity tomography (ERT) surveys. J Appl Geophys 146:103--119

Uhlemann S, Chambers J, Wilkinson P et al (2017) Four-dimensional imaging of moisture dynamics during landslide reactivation. J Geophys Res Earth Surf 122(1):398--418

Weller A, Slater L (2012) Salinity dependence of complex conductivity of unconsolidated and consolidated materials: comparisons with electrical double layer models. Geophysics 77(5):D185--D198

Weller A, Slater L, Nordsiek S (2013) On the relationship between induced polarization and surface conductivity: implications for petrophysical interpretation of electrical measurements. Geophysics 78(5):D315--D325

Weller A, Slater L, Binley A et al (2015) Permeability prediction based on induced polarization: insights from measurements on sandstone and unconsolidated samples spanning a wide permeability range. Geophysics 80(2):D161--D173

Whiteley J, Watlet A, Uhlemann S et al (2021) Rapid characterisation of landslide heterogeneity using unsupervised classification of electrical resistivity and seismic refraction surveys. Eng Geol 290:106189

Wilkinson P, Chambers J, Uhlemann S et al (2016) Reconstruction of landslide movements by inversion of 4-D electrical resistivity tomography monitoring data. Geophys Res Lett 43(3):1166--1174

Xu S, Sirieix C, Riss J et al (2017) A clustering approach applied to time-lapse ERT interpretation -- case study of Lascaux cave. J Appl Geophys 144:115--124

Yao W, Li C, Zhan H et al (2019) Time-dependent slope stability during intense rainfall with stratified soil water content. Bull Eng Geol Env 78(7):4805--4819

Zhang Z, Weller A, Kruschwitz S et al (2019) Effect of evaporative drying on complex conductivity spectra of sandstones. Geophysics 84(1):MR61--MR72

Zisser N, Kemna A, Nover G (2010) Relationship between low-frequency electrical properties and hydraulic permeability of low-permeability sandstones. Geophysics 75(3):E131--E141

\endgroup

\end{document}